# Integration of Fractional Order Black-Scholes Merton with Neural Network

Sarit Maitra
*Alliance School of Business*
*Alliance University*
Bengaluru, India
sarit.maitra@alliance.edu.in

Vivek Mishra
*School of Applied Mathematics*
*Alliance University*
Bengaluru, India
vivek.mishra@alliance.edu.in

Goutam Kr. Kundu
*Alliance School of Business*
*Alliance University*
Bengaluru, India
goutam.kundu@alliance.edu.in

Kapil Arora
*Alliance School of Business*
*Alliance University*
Bengaluru, India
kapil.arora@alliance.edu.in

*Abstract*— This study presents a novel approach to enhance option pricing accuracy by introducing the Fractional Order Black-Scholes-Merton (FOBSM) model. FOBSM combines elements of the traditional Black-Scholes-Merton (BSM) model with the flexibility of neural networks (NN). American options pose unique pricing challenges due to free boundary difficulties. On the other hand, traditional models like BSM struggle to accurately represent market pricing. The challenge is to develop a pricing model that better captures the tail behavior, memory effects, volatility clustering, long-term dependencies, and skewness inherent in financial data, while simultaneously utilizing the theoretical underpinnings of BSM and fractional calculus. The gap arises from the absence of a comprehensive framework that integrates fractional calculus and neural networks to enhance option pricing accuracy in complex diffusion dynamics scenarios. Since FOBSM captures memory characteristics in sequential data, it is better at simulating real-world systems than integer-order models. Findings reveal that in complex diffusion dynamics, this hybridization approach in option pricing improves the accuracy of price predictions.

## I. Introduction

American options offer greater flexibility and profit potential than European options, but they are valued differently. European options use mathematical formulas for pricing, while American options have an early exercise feature. The presence of early exercise options creates free boundary difficulties in the pricing process, making it more challenging than European options pricing. This research integrates fractional calculus and neural networks into option pricing and presents a theoretical framework to address this challenge. Fractional calculus (FC), which uses derivatives and integrals of any non-integer order, has gained popularity for its capacity to simulate memory-like systems ([1], [2]). Model based on fractional calculus account for volatility clustering, long-term dependencies, tail behavior, and skewness, making them better for options with longer maturities or market stress.

Black and Scholes [3] introduced the fundamental equation in 1973, and Merton later modified it, earning them the Nobel Memorial Prize. This is known as the Black-Scholes-Merton (BSM) model which provides a solid theoretical framework. It has limitations when it comes to accurately reflecting actual market pricing. Despite this, the model is a helpful tool for evaluating options. However, a mathematically simpler binomial model was proposed [4] which is intuitive and more frequently used in practice for discrete price data. Our work pertains to continuous data, and it employs the BSM model as the foundation for option pricing. While recognizing the BSM model's theoretical significance, our study aims to enhance its applicability and precision by incorporating fractional calculus and neural networks to address the challenge posed by complex market dynamics and discrepancies between theoretical and market-based option pricing.

The lognormal diffusion assumption is a fundamental assumption in the Black-Scholes-Merton (BSM) option pricing model, which plays a key role in the model's formulation. This assumption is incompatible with market prices, which led to the BSM model being modified over the years to enhance its performance. Examples of these modifications include jump diffusion models ([5], [6]), stochastic volatility models ([7], [8]), and others. We introduce fractional order Black-Scholes-Merton (FOBSM), which is an extension of the classical BSM that incorporates fractional calculus. A combination of closed-form and numerical solutions to address mathematical problems has been examined by researchers [9]. Their work has inspired the solution of FOBSM, which allows for the analysis of implied volatility and option prices' non-local behavior. FOBSM uses neural networks' flexibility and adaptability to improve option pricing accuracy in situations where traditional models may fall short. The hybridization of FOBSM and NN in options increases the accuracy of option price predictions, particularly in situations with complicated diffusion dynamics. The findings are divided into two categories: providing option values using more powerful numerical and analytic approaches and developing new pricing models that accurately reflect the financial market.

The major contributions to this work are:

1) A novel approach to option pricing that leverages the strengths of both the FOBSM and NNs (Neural Networks) by integrating both models.
2) Improved accuracy in predicting option prices, especially in scenarios with complex diffusion dynamics, which refers to the movement or behavior of an underlying asset's price over time.

### A. Black Scholes model with default risk

The variables used by BSM are the volatility, underlying asset price, option strike price, expiration time, and risk-free interest rate.

Eq. (1) and (2) formulate the put and the call options:

$$Call = S_0 N(d_1) - N(d_2) K e^{-rT} \qquad (1)$$

$$Put = N(-d_2) K e^{-rT} - N(-d_1) S_0 \qquad (2)$$

where, $d_1 = \left( ln(S/K) + (r + \frac{\sigma^2}{2})T \right) / \sigma\sqrt{T}$, $d_2 = d_1 - \sigma\sqrt{T}$ and $S$ = underlying asset prices corresponding to the observed option prices at time $t$, $N$ = normal distribution, $K$ = strike price, $t$ = time to maturity, $\sigma$ = volatility (standard deviation of log return), $r$ = risk-free interest rate, $T$ = time to expiration for each option.

The value of a call option is divided into two parts:

- the amount paid when exercising the option ($N(d_2) K e^{-rT}$)
- the amount received when selling the underlying ($N(d_1) S_t$).

The strike price (K) undergoes a discounting process using the standard discounting factor ($e^{-rT}$), where $e$ represents the mathematical constant, and $r$ signifies the risk-free interest rate applicable for investing funds over a specified period $T$. The determination of this risk-free rate ($r$) can be achieved through a range of established methodologies. The cumulative distribution function (CDF) of a normal distribution serves as a crucial component in financial modeling. It provides the probability that a random variable will assume a value less than or equal to the input provided to the CDF. This aids in assessing the likelihood of various financial events and outcomes. In the context of option pricing, the variables ($d_1$ and $d_2$) play a pivotal role. As these variables increase in value, so does the corresponding cumulative distribution function $N(d_1)$, and $N(d_2)$ respectively. This, in turn, directly impacts the valuation of the option. Additionally, as the underlying asset's price ($S_t$) experiences an increase, the option's intrinsic value also rises, thereby resulting in potential gains associated with the exercise of the option. These dynamics underscore the critical relationship between the various parameters involved in option pricing and the potential financial outcomes.

Volatility (σ) is a crucial factor in option pricing. The presence of volatility in both the numerator and denominator of the BSM formula reflects its importance in determining the option's price. The square of volatility in the numerator is associated with the volatility's effect on the stock's expected drift over time. Higher volatility leads to larger potential price movements, which results in a higher expected return and therefore a higher option price. BSM model assumes constant volatility over time, which may not always be an accurate representation of real-world markets. The implied volatility, which represents market expectations of future volatility, can vary across different strike prices of options with the same expiration date. This leads to the observation of a volatility skew when implied volatilities are graphed against strike prices. These shortcomings of the BSM have led some researchers to place more importance on historical volatility as opposed to implied volatility (e.g., [10]).

The underlying lognormal distribution in BSM is frequently used to simplify the behavior of stock prices over time. Researchers have solved the problem using various analytical and numerical methods (e.g., [11], [12], [13], [3], etc.). Many models now in use, however, make rigid assumptions, such as ideal markets, constant risk-free rates and volatility, a lognormal distribution of share price dynamics, the absence of dividends, and continuous delta hedging.

## II. PROBLEM FORMULATION

Our objective is to present an integrated model that combines the FOBSM with a NN to predict option prices more reliably.

Considering the variables:

- $C$ = Option price as a function of the underlying asset price S and time to expiry t,
- $C_{obs}$ = actual prices for a set of options.
- $S$ = underlying asset prices corresponding to the observed option prices at time t.
- $\theta$ = Fractional order parameters with respect to underline price characterizing the fractional diffusion process in the FOBSM.
- $\alpha$ = Fractional order parameter with time
- $N$ = Neural Network with parameters
- $W$ = weights and biases.
- $C_{pred}$ = predicted option prices generated by the integrated model.
- $TC$: Transaction costs associated with buying or

Because of TC, investors are deterred from trading when market impulses are prevalent. Thus, TCs are an important consideration in financial markets and can lead to non-linear dynamic adjustments in option prices [14]. Eq. (3) is a stochastic differential equation describes how the variable S normally moves in a geometric Brownian motion:

$$dS(t) = rS(t)dt + \sigma S(t)dW(t) \qquad (3)$$

where, $r$ = risk free interest, $\sigma$ = volatility of the stock returns, $dW$ = Wiener process (Brownian motion). The assumption of BSM's lognormal diffusion can be expressed by Eq. (3) as well.

We introduce a set of unknown parameters $\Phi = (\Phi_1, \Phi_2, \ldots, \Phi_k)$ that influence the functions $\mu = \mu_\Phi$ and $\sigma = \sigma_\Phi$ in the stochastic model. These parameters are estimated to maximize the log-likelihood function as displayed in $d_1$ and $d_2$ in Equations (1) & (2) which help determine the best-fit values based on observed data.

Thus, our goal is to develop a model that considers transaction costs, the stochastic nature of asset price movements, and unknown parameters to make predictions that better reflect real-world market dynamics, including skewness, kurtosis, and volatility skew.

## III. FRACTIONAL ORDER BLACK-SCHOLES MODEL (FOBSM)

To achieve the goal, we introduce a time-fractional BSM by employing Riemann-Liouville fractional derivatives within the framework of partial differential equations (PDEs). The Riemann-Liouville mathematical tool facilitates the computation of fractional derivatives. It plays a vital role in capturing the intricate dynamics of financial data, particularly when dealing with long memory, fractal behavior, or non-Gaussian distributions.

TABLE 1. PSEUDOCODE (RIEMANN-LIOUVILLE FRACTIONAL DERIVATIVES)

```
function fractional_derivative(C, alpha, delta_t):
    n = length(C)
    result = array of zeros with the same length as C

    for i from 1 to n - 1:
        integral_sum = 0.0
        for k from 0 to i - 1:
            integral_sum = integral_sum + ((i - k) ^ (-alpha)) * C[k]
        result[i] = (1 / gamma(1 - alpha)) * integral_sum * (delta_t ^ (-alpha))

    return result
```

Following the introduction of the traditional BSM, PDEs theory and methodology started to gain popularity for the research of option pricing issues. Several researchers have emphasized the study of fractional differential equations (e.g., [15]; [16]; etc.). Eq. (4) presents the option price using the BSM equation.

$$C_{FBM}(S, K, T, r, \sigma) = Ke^{-rT}N(-d_2) - SN(-d_1) \quad (4)$$

FOBSM is given by Eq. (5), which extends the partial differential equations by applying fractional derivatives with both 1st order and 2nd order derivatives:

$$\frac{\partial^\alpha C}{\partial t^\alpha} + rS\frac{\partial^\theta C}{\partial S^\theta} + \frac{1}{2}\sigma^2 S^2 \frac{\partial^{2\theta} C}{\partial S^{2\theta}} - rC - TC(t) = 0 \quad (5)$$

where, $0 < \alpha < 1, S \geq 0, t \geq 0$ and $t \in [0, T]$, where $0 < \theta < 1$

Eq. (5) is the continuous form, where $\partial^\alpha C / \partial t^\alpha$ = fractional time derivative of the option price C with respect to time t, where α = order of the time fractional derivative. This portion of our work is based on the research work [17], which derived a numerical scheme of second order in space. Eq. 6 displays the fractional time derivative.

$$\frac{\partial^\alpha C}{\partial t^\alpha} = \frac{1}{\Gamma(1-\alpha)} \int_0^t (t-\tau)^{-\alpha} \frac{\partial C}{\partial \tau} d\tau \quad (6)$$

The FOBSM equation (Eq. 5) with Riemann-Liouville fractional derivatives becomes:

$$\frac{1}{\Gamma(1-\alpha)} \int_0^t (t-\tau)^{-\alpha} \frac{\partial C}{\partial \tau} d\tau + rS\frac{\partial^\theta C}{\partial S^\theta} + \frac{1}{2}\sigma^2 S^2 \frac{\partial^{2\theta} C}{\partial S^{2\theta}} - rC - TC(t) = 0 \quad (7)$$

where, $0 < \alpha < 1, S \geq 0, t \geq 0$ and $t \in [0, T], 0 < \theta < 1$

Eq. 8 displays the discretize form of this using finite difference:

$$\frac{\partial^\alpha C}{\partial t^\alpha} \approx \frac{1}{\Gamma(1-\alpha)} \sum_{k=0}^{i-1} (t_i - t_k)^{-\alpha} (\Delta C_k) u \quad (8)$$

Where, t = current time step, $t_i$ = current time, $\Delta t$ = time step size, $\Delta C_k$ = change in C from time $t_k$ to $t_{k+1}$.

The fractional derivative of order θ approximated as:

$$\frac{\partial^\theta C}{(\partial S)^\theta} \approx D_s^\theta C_{i,j} \quad (9)$$

where $D_s^\theta$ is fractional difference operator.

Contrary to integer-order differential operators, which are local, fractional-order differential operators are nonlocal, allowing a system's next state to depend on both its present state and its past state ([18], [19]). When we add fractional derivatives to the Black-Scholes-Merton model, we are using fractional-order differential operators.

*A. Stationarity assumptions*

The stationarity assumption is a fundamental concept in time series analysis and stochastic processes. In the context of financial modeling, including the BSM model and its extensions, stationarity assumptions are important considerations.

*1) Fractional BSM Model and Stationarity:*

Fractional calculus deals with derivatives and integrals of non-integer orders. It extends the traditional concepts of differentiation and integration to include fractional or non-integer orders. Though fractional calculus allows for the incorporation of non-standard diffusion phenomena, including long memory and volatility clustering; however, the degree of non-stationarity introduced depends on the choice of fractional orders and the modeling assumptions, so the level of stationarity vs. non-stationarity can vary in practice. In this case, the selection of fractional orders (α and θ) in the fractional calculus equations is a crucial decision. Different values of these can lead to varying degrees of non-stationarity and memory effects in the model. We need to calibrate these parameters based on the specific characteristics of the data. Thus, when introducing fractional derivatives to the BSM model, the assumption of stationarity becomes more flexible.

## IV. NEURAL NETWORK INTEGRATION

The ability of neural networks to approximate the relationship between multiple Black-Scholes parameters and the ultimate option price has been established by researchers ([20], [21]). The growing utilization of neural networks as function approximators [22] and numerical solvers has opened new horizons in the fields of machine learning and numerical analysis ([23], [24]). Building upon these foundational contributions, a comprehensive literature review was presented [25] which reveals that neural networks have been employed in option pricing since the early 1990s.

Neural networks, particularly those employing the Universal Approximation-Rectified Linear Unit (ReLU), possess the capability to approximate any continuous function within bounded subsets of real numbers [26]. Furthermore, empirical evidence suggests that neural networks often outperform the Black-Scholes model in many scenarios [27]. Notably, neural networks tend to excel in stable market conditions, whereas the Black-Scholes model exhibits superiority in more volatile market environments [28]. These findings have served as compelling motivation for the integration of neural networks into our research study.

Fig. 2 displays the convergence of the mean squared error (MSE) for the integrated model using a different optimizer (Adam, SGD, RMSprop). During training, each optimizer adjusts the model's weights to reduce the MSE between the predicted and actual option prices.

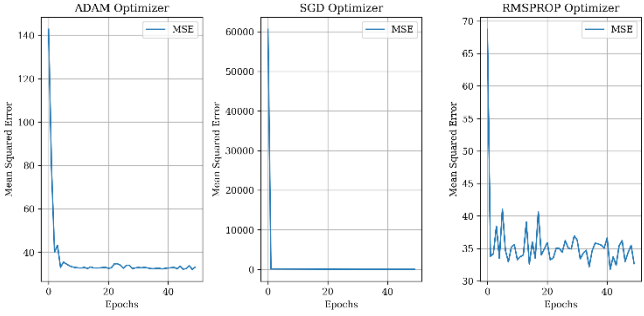

Fig. 1.  Comparison of different optimizers.

From the plots, we can see ADAM consistently results in a lower MSE and converges more quickly; it can be considered more effective for this specific problem. However, we have used a custom loss function, which combines both mean squared error (MSE) loss and a BSM-based loss.

## V. MODEL ARCHITECTURE AND LOSS

We simulated realistic financial scenarios to generate synthetic with the key parameters S, K, T, r, σ and θ. These parameters vary within predefined ranges to create diverse data points. The fractional BSM introduces two fractional orders: α (for time) and θ (for spatial derivatives).

### A. Fractional Derivative Implementation

We implement fractional derivatives using the Riemann-Liouville fractional derivative numerical approximation. It iterates over points in the time domain and computes the left Riemann-Liouville fractional derivative for each point based on the generated data.

### B. Discretization Process

To calculate the fractional derivative numerically, we discretized the integral in the Riemann-Liouville formula. For each data point, the code computes option prices using both the fractional Black-Scholes formula and the traditional Black-Scholes formula. The fractional Black-Scholes formula mitigates issues related to singularity and avoids division by zero by adding a small constant (ϵ). Three key features for each data point were extracted:

- The last value of the option price
- The last value of the fractional time derivative
- The last value of the fractional spatial derivative

Normalization is performed on these features by subtracting the mean and dividing by the standard deviation to ensure that features have similar scales.

We design and construct a simple feedforward neural network (NN) to test our hypothesis. The architecture encompasses an initial input layer comprised of two neurons. These two neurons are dedicated to encoding the essential information regarding the normalized fractional time derivative and fractional spatial derivative, both of which are integral components of this model. Within the NN, there are two subsequent hidden layers, each of considerable depth, housing 64 neurons in each layer. These hidden layers are pivotal in the neural network's ability to capture intricate patterns and relationships within the data. To facilitate the activation of these hidden neurons, Rectified Linear Unit (ReLU) activation functions have been employed. The predictive capability is realized through the presence of an output layer with a singular neuron. This singular neuron serves as the neural network's interface for generating predictions related to the anticipated option prices, thereby encapsulating the core objective of the study.

### C. Loss function

The NN compiled using the Adam optimizer and mean squared error (MSE) loss function.

$$\text{MSE} = \frac{1}{n}\sum_{i=1}^{n}(\hat{y}_i - y_i)^2 \qquad (10)$$

The network trained using the training data for 500 epochs. The batch size is set to 32, and 20% of the training data is used as a validation set. Early stopping is employed to prevent overfitting, monitoring the validation loss for improvement. Fig. 2 & 3 display the training – validation loss and actual vs predicted option pricing, respectively.

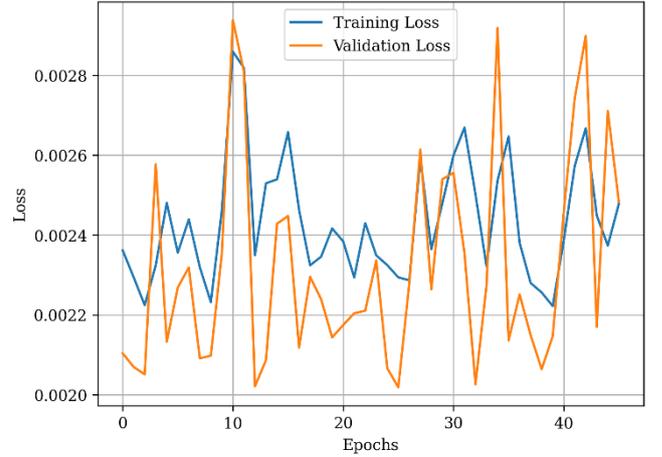

Fig. 2.  Learning curve.

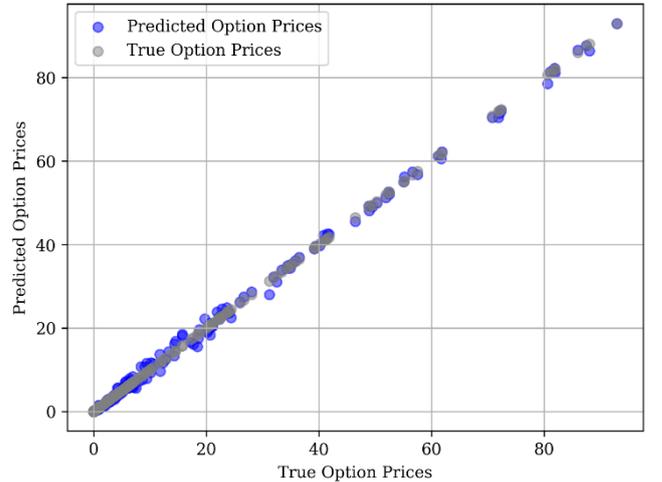

Fig. 3. True vs Predicted option prices.

To this end, this article presents a novel approach to option pricing that combines fractional calculus, numerical methods, adaptive techniques, and neural networks. Based on the theoretical foundation and experiments with synthetic data, this approach is likely to improve the accuracy of predicting option prices by accounting for memory, long-term dependencies, and complex behaviors. It also adds a data-driven element, capturing intricate financial data patterns and offering an alternative perspective on option pricing.

The model is trained and tested on synthetic data. While testing on synthetic data can help assess how well the model performs under controlled conditions, it is essential to validate the model's performance on real financial data which is the future direction beyond the scope of this work.

TABLE 2. COMPUTER CODE

```
"""fractional derivative using Riemann-Liouville method"""
def fractional_derivative(C, alpha, delta_t):
    n = len(C)
    result = np.zeros_like(C)

    for i in range(1, n):
        integral_sum = 0.0
        for k in range(i):
            integral_sum += ((i - k) ** (-alpha)) * C[k]
        result[i] = (1 / gamma(1 - alpha)) * integral_sum * (delta_t ** (-alpha))
    return result

"""synthetic data"""
num_data_points = 1000

S = np.random.uniform(50, 150, num_data_points)
K = np.random.uniform(50, 150, num_data_points)
T = np.random.uniform(0.1, 1.0, num_data_points)
r = np.random.uniform(0.01, 0.05, num_data_points)
sigma = np.random.uniform(0.1, 0.5, num_data_points)

"""option prices using a time and price fractional BSM"""
alpha_price = 0.7
alpha_time = 0.5
option_prices = np.zeros((num_data_points,))
fractional_price_derivatives = np.zeros((num_data_points,))
fractional_time_derivatives = np.zeros((num_data_points,))

for i in range(num_data_points):
    t = np.linspace(0, T[i], 1000)
    delta_t = t[1] - t[0]
    S0 = S[i]
    K0 = K[i]
    r0 = r[i]
    sigma0 = sigma[i]
    epsilon = 1e-10
    if sigma0 < epsilon or t[-1] < epsilon:
        continue

    """option price using the BSM"""
    d1 = (np.log(S0 / (K0 + epsilon)) + (r0 + (sigma0 ** 2) / 2) * t) / (sigma0 * np.sqrt(t))
    d2 = d1 - sigma0 * np.sqrt(t)

    option_price = (
        K0 * np.exp(-r0 * t) * norm.cdf(-d2) - S0 * norm.cdf(-d1))

    """fractional price derivative for this data point"""
    fractional_price_deriv = fractional_derivative(option_price, alpha_price, delta_t)

    """fractional time derivative for this data point"""
    fractional_time_deriv = fractional_derivative(option_price, alpha_time, delta_t)

    option_prices[i] = option_price[-1]
    fractional_price_derivatives[i] = fractional_price_deriv[-1]
    fractional_time_derivatives[i] = fractional_time_deriv[-1]

"""Normalize the data"""
mean_option_prices = option_prices.mean()
std_option_prices = option_prices.std()
mean_price_derivatives = fractional_price_derivatives.mean()
std_price_derivatives = fractional_price_derivatives.std()
mean_time_derivatives = fractional_time_derivatives.mean()
std_time_derivatives = fractional_time_derivatives.std()

normalized_option_prices = (option_prices - mean_option_prices) / std_option_prices
normalized_price_derivatives = (fractional_price_derivatives - mean_price_derivatives) / std_price_derivatives
normalized_time_derivatives = (fractional_time_derivatives - mean_time_derivatives) / std_time_derivatives

"""neural network model"""
def build_neural_network(input_dim):
    model = tf.keras.Sequential([
        tf.keras.layers.Input(shape=(input_dim,)),
        tf.keras.layers.Dense(64, activation='relu'),
        tf.keras.layers.Dense(64, activation='relu'),
        tf.keras.layers.Dense(1)
    ])
    return model

X = np.column_stack((normalized_price_derivatives, normalized_time_derivatives))

neural_network = build_neural_network(input_dim=2)
neural_network.compile(optimizer='adam', loss='mse')

train_size = int(0.8 * len(X))
X_train, X_test = X[:train_size], X[train_size:]
y_train, y_test = normalized_option_prices[:train_size], normalized_option_prices[train_size:]

early_stopping = tf.keras.callbacks.EarlyStopping(monitor='val_loss', patience=20, restore_best_weights=True)

neural_network.fit(X_train, y_train, epochs=500, batch_size=32, validation_split=0.2, callbacks=[early_stopping])
mse = neural_network.evaluate(X_test, y_test)
predicted_option_prices = neural_network.predict(X_test)
denormalized_predictions = (predicted_option_prices * std_option_prices) + mean_option_prices
```

## VI. LIMITATIONS & FUTURE DIRECTION

While the work introduces a novel and promising approach to option pricing, it should be seen as a starting point for further research and refinement. It is theoretical in nature and focuses on integrating mathematical and computational techniques. The model's performance needs to be validated against real-world option pricing scenarios and historical data to assess accuracy and robustness. The integration involves tuning multiple parameters, including the trade-off parameter ($\lambda$) and neural network architecture and optimization parameters. Finding the optimal combination of these parameters can be challenging and time-consuming. Practitioners often use cross-validation, grid search, random search, or Bayesian optimization to systematically explore the parameter space and identify a reasonable set of parameters to solve these difficulties. FOBSM's accuracy is required to be empirically evaluated using historical data. The financial industry is known for its conservative approach to adopting new models and technologies. Investigating how well the proposed model can be integrated into existing financial systems and regulatory frameworks is crucial for its real-world adoption.

## VII. CONCLUSION

This study marks a substantial advancement in the pursuit of greater precision in option pricing models. Through the fusion of mathematical rigor and computational prowess, our objective has been to diminish the divide between theoretical models and the intricate realities of financial markets. It is imperative to acknowledge that our scholarly voyage remains an ongoing endeavor. The trajectory toward refining option pricing is a dynamic and ever-evolving domain within research.